\documentstyle[epsfig]{elsart}              

\newcommand{\bmf}[1]{\mbox{\boldmath $#1$}}
\newcommand{\bfx}{\mbox{\scriptsize \boldmath $x$}}
 
\def\gsim{\hbox{\lower3pt\vbox{\baselineskip=4pt \lineskiplimit=0pt \kern2pt 
          \hbox{$>$}\hbox{$\sim$}}}}                                                                                
                                                                               
\begin{document}
\begin{frontmatter}

                                                                              
\title{Dynamic relaxation of SU(2) lattice gauge theory in
(3+1) dimensions} 

\author{A.~Jaster}
\address{Universit\"{a}t - GH Siegen, D-57068 Siegen, Germany}
\date{\today}
                                                                                

\maketitle
\begin{abstract}

We investigate the  dynamic relaxation for 
SU(2) gauge theory at finite temperatures in $(3+1)$ dimensions.
Using the Hybrid Monte Carlo algorithm, we examine the time 
dependence of the system in the short-time regime.
Starting from the ordered state, the
critical exponents $\beta$, $\nu$ and $z$ are calculated
from the power law behaviour of the Polyakov loop and the cumulant at 
or near the critical point. The results for the static exponents 
are in agreement with those obtained from simulations in equilibrium
and those of the three-dimensional Ising model. The value for the dynamic 
critical exponent was determined with $z=2.0(1)$. 
\end{abstract}

\\
\endkeyword

\end{frontmatter}

\section{Introduction}

Traditionally, it was believed that universal scaling behaviour exists
only in the long-time regime. However, recently
Janssen, Schaub and Schmittmann \cite{JASCSC} showed that far
from equilibrium, in a macroscopic short-time regime of the dynamic evolution,
there already emerges universal scaling behaviour in the O($N$) vector model.
They considered the relaxation process of a system  quenched from 
a disordered state to the critical point and evolving with
dynamics of model A (non-conserved order parameter and non-conserved 
energy \cite{HOHHAL}) and found that
the magnetization undergoes an initial increase of the form $M(t) \sim 
t^\theta $, where $\theta$ is a new dynamic exponent and $t$ denotes the time. 
This prediction was
supported by a number of Monte Carlo (MC) investigations not only for 
O($N$) vector models, but also for several other systems with a second-order
or a Kosterlitz-Thouless phase transition \cite{ZHENG}.
These simulations offer also a possibility to determine the conventional
(static and dynamic) critical exponents \cite{LISCZH1,LISCZH2,SCHZHE} 
as well as the critical point \cite{SCHZHE}.
This may eliminate critical slowing down, since the measurements
are performed in the early part of the evolution.

In short-time critical dynamics exist two different time scales, the microscopic
time scale $t_{\mathrm{mic}}$ and the macroscopic time scale 
$t_{\mathrm{mac}}$.  
Universal behaviour emerges only after $t_{\mathrm{mic}}$,
the time during which the non-universal short wave behaviour is swept away.
$t_{\mathrm{mic}}$ is
independent of the linear space extension $L$ and it is in general small compared to
the macroscopic time scale, which is proportional  $ L^z $.
After a time period of $t_{\mathrm{mac}}$ the correlation length is of the order
of $L$ and the system crosses over to the long-time universal behaviour.

First numerical simulations of the short-time dynamical relaxation
at criticality started from
a disordered initial state. However, dynamical scaling exists also for
an ordered initial state. This is supported by a variety of MC
investigations \cite{ZHENG,JAMASCZH,JASTER0}, but
no analytical calculations exist for this situation.
All static  exponents as well as the critical point 
and the correlation length in the high temperature phase \cite{JASTER} 
can be also obtained with an ordered initial state.

Systematic numerical simulations of the short-time critical dynamics have been
carried out mainly in two-dimensional systems \cite{ZHENG,LUSCZH,YILUSCZH,JASTER0}. 
Also, a first approach to lattice 
gauge theory was done for SU(2) gauge fields in $(2+1)$ dimensions
\cite{OKSCZH}. The results strongly support that there exists a universal short-time
scaling behaviour for the dynamic SU(2) lattice gauge theory.
In this paper we  investigate SU(2) gauge theory 
in $(3+1)$ dimensions. In contrast to the previous study in $(2+1)$ dimensions
we start the dynamic evolution from an initial ordered state, i.e.\ with a
magnetization of one at $t=0$ ($m_0=1$). For the determination of the critical exponents,
the dynamic relaxation process starting from an ordered state has been proven
advantageous over that from an unordered initial state. This is a consequence 
of less prominent fluctuations for an ordered initial
state. However, the new critical exponent $\theta$
can not been calculated if we start with $m_0=1$.

The dynamics of our system is given by the Hybrid Monte Carlo (HMC) algorithm
\cite{DUKEPERO}.
Up to now mainly local algorithms such as the Metropolis or the heatbath algorithms
have been used for the dynamic evolution. This is the first time that the HMC
algorithm is studied in critical dynamic relaxation. 
The motivation for using the HMC algorithm is not only to enlarge the knowledge
of short-time dynamics, but also the possibility to include fermions in
future simulations.
We examine the short-time behaviour of the order parameter and the cumulant
at and in vicinity of the critical point. From the power law behaviour 
we extract the critical exponents $\beta$, $\eta$ and $z$.
The results are compared with those of simulations in equilibrium \cite{SU2EXP}
and with simulations of the
three-dimensional Ising model \cite{JAMASCZH}, since it is expected that both
models are in the same universality class \cite{SVEYAF}.

In the next section we sketch the scaling analysis of the 
short-time critical dynamics. Sections 3 and 4 briefly describe the model 
and the updating algorithm.
Numerical results are presented in Sec.\ 5. 
The last section contains the summary and
conclusions.

\section{Scaling relations}

Using renormalization group methods, Janssen et al.\ have shown that even
in macroscopically early stages of a relaxation process O($N$)
vector models  display universal
behaviour \cite{JASCSC}. They studied 
a system initially in a disordered state with vanishing or small
magnetization ($m_0 \,\gsim\, 0$), suddenly
quenched to the critical point and evolving with dynamics of model A,
and derived the dynamic scaling form 
\begin{equation}
M^{(k)}(t,\tau,L,m_0) = b^{-k \beta / \nu} M^{(k)} (b^{-z}t,b^{1/\nu}\tau,
b^{-1}L, b^{x_0}m_0)\, .
\end{equation}
$M^{(k)}$ denotes the $k$th moment of the magnetization,
$t$ is the MC time of the dynamic relaxation, $\tau$ is the 
reduced coupling constant, $b$ indicates a spatial rescaling factor
and $x_0$ is a new independent exponent. For a sufficiently large lattice
and small initial magnetization 
$m_0 \, t^{x_0/z}$ this leads to
\begin{equation}
M(t) \sim m_0 \, t^{\theta} \, , \ \ \theta=(x_0 -\beta/\nu)/z
\end{equation}
at the critical point $\tau=0$.

Another important process is the dynamic relaxation from a completely 
ordered state. For an initial magnetization exactly at its fixed point
$m_0=1$, a scaling form
\begin{equation}
\label{scalingform}
M^{(k)}(t,\tau,L) = b^{-k \beta / \nu} M^{(k)} (b^{-z}t,b^{1/\nu}\tau,
b^{-1}L)
\end{equation}
is expected. The scaling form 
(\ref{scalingform}) looks the same as the dynamic scaling form in the long-time
regime, however, it is now assumed already valid in the macroscopic short-time
regime. The validity of the scaling form (\ref{scalingform}) 
in the short-time regime was verified with MC simulations 
for a number different systems \cite{ZHENG}.

Taking $b=t^{1/z}$ for the spatial rescaling factor in Eq.\ 
(\ref{scalingform}) with $k=1$ leads 
for the magnetization to a power  law  behaviour 
\begin{equation}
\label{eqplm}
M(t) \sim t^{-c_1} \, , \ \ c_1=\frac{\beta}{\nu z}
\end{equation}
at the critical point $\tau=0$, if $L$ is sufficiently large.
For non-zero values of $\tau$, the power law behaviour
will be modified by the scaling function $M(1,t^{1/\nu z} \tau)$.
This can be used for a determination of the critical point. Also,
the critical exponent $1/(\nu z)$ \cite{SCHZHE} can be measured
by taking the derivative with respect to $\tau$
\begin{equation}
\partial_{\tau} \ln M(t,\tau) |_{\tau=0} \sim t^{c_{l1}} \, ,
\ \ c_{l1}=\frac{1}{\nu z} \, ,
\end{equation}
while the dynamic critical exponent $z$ can be determined from the
behaviour of the cumulant 
\begin{equation}
U(t)=\frac{M^{(2)}}{(M)^2}-1 \, .
\end{equation}
Finite-size scaling shows that
\begin{equation}
U(t) \sim t^{c_U} \, , \ \ c_U=\frac{d}{z} \ ,
\end{equation}
where $d$ denotes the spatial dimension.
Thus, the short-time behaviour of the dynamic relaxation starting from a 
completely ordered state is sufficient to determine all the critical exponents
$\beta$, $\nu$ and $z$ as well as the critical point.
These measurements are usually better in quality than those
starting from a disordered state.

\section{SU(2) gauge theory at finite temperatures}

The Wilson action for SU(2) gauge theory is given by
\begin{equation}
S= \frac{4}{g^2}
 \sum_{\mathrm{P}}  \left ( 1 - \frac{1}{2} \,{\mathrm{Tr}}\, 
U_{\mathrm{P}} \right ) \, ,
\end{equation}
where $U_{\mathrm{P}}$ represents the usual plaquette term on the lattice.
The number of lattice points in the space direction is $L_{\mathrm{s}}$
and in the time direction $L_{\mathrm{t}}$. Thus, the volume and temperature are 
given by $V= {L_{\mathrm{s}}}^3 L_{\mathrm{t}}$ and $T=1/L_{\mathrm{t}}$,
if we fix the lattice spacing $a$ to unity. A point on the lattice has integer 
coordinates $x=(x_0,\bmf{x})=(x_0,x_1,x_2,x_3)$, which are in the range
$0 \le x_0 < L_{\mathrm{t}}$, $0 \le x_i < L_{\mathrm{s}}$ ($i=1,2,3$).
A gauge field  $U_{x,\mu}$ is assigned to the link pointing from point
$x$ to point $(x+\mu)$, where $\mu =0,1,2,3$ designates the four
forward directions in space-time.

The order parameter (magnetization) of the system at some MC time $t$ 
is the expectation 
value of the Polyakov loop
\begin{equation}
M(t)= \frac{1}{{L_{\mathrm{s}}}^3} \sum_{\bfx} \left \langle 
L_{\bfx}(t) \right \rangle \, ,
\end{equation}
which is defined  as the trace of ordered products of gauge field variables
\begin{equation}
L_{\bfx}(t)= \frac{1}{2} \, {\mathrm{Tr}} \prod_{x_0=0}^{L_{\mathrm{t}}-1} 
U_{(x_0,{\bfx}),0}(t) \, .
\end{equation}
The average is taken over independent measurements, i.e.\ independent
random numbers.
The deconfining phase transition of this model is of second order.
The critical point for $L_{\mathrm{t}}=4$ was determined
with $4/{g_{\mathrm{c}}}^2=2.2989(1)$ for infinite large space extensions
\cite{SU2EXP}.

\section{The HMC algorithm}

Let us briefly sketch the HMC algorithm. In ordinary Metropolis or heatbath
updating algorithms  the new configuration is generated 
by sweeping over the whole 
system and changing locally the field variables $q$. In case of the HMC
algorithm one uses molecular dynamics to generate the new configurations.
One starts by introducing an additional, fictitious so-called 
molecular dynamics time
$t'$ and corresponding momenta $p$. 
The initial  conjugate momenta  $p_i$ are generated from a Gaussian
distribution of unit variance and zero mean. The fictitious time 
evolution of the fields and the momenta is now given by the following set of 
coupled first-order differential equations:
\begin{equation}
\label{Eqsmot0}
\dot{p}_i=- \frac{\partial {\cal H}}{\partial q_i} \, , \ \ 
\dot{q}_i=p_i \, ,
\end{equation}
where the Hamiltonian is given by ${\cal H}=\sum_i {p_i}^2/2 + S[q]$.
The time derivates are to be understood with respect to the fictitious time $t'$.
The numerical integration of Eqs.\ (\ref{Eqsmot0}) is performed by using a
discretized version. In practice one uses a leap-frog integration scheme,
using $N_{\mathrm{MD}}$ integration steps of size $\Delta t'$ in order
to integrate from fictitious time $0$ to some value $t'=N_{\mathrm{MD}} 
\, \Delta t'$. 
The endpoint of these trajectories are
considered as a trial new configuration, which is accepted or rejected 
according to  the general Metropolis acceptance probability. The HMC algorithm is
exact\footnote{The lack of reversibility coming from round-off errors
in the numerical integration are discussed in Ref.\ \cite{LIJAJA}.}, 
i.e.\ systematic errors arising from 
finite time steps in the molecular dynamics are avoided by the accept/reject
step. The algorithm is also ergodic due to the stochastic update of the
initial momenta and fulfills the detailed balance condition, 
because of the reversibility
of the leap-frog integration. 

For SU(2) lattice gauge theory the equations of motion are  
\begin{equation}
\label{Eqsmot}
{\mathrm{i}} \dot{H}_{x,\mu} = -  \frac{4}{g^2} \left (
U_{x,\mu} V_{x,\mu} - \mbox{h.c.} 
\right ) \, , \ \
\dot{U}_{x,\mu} = {\mathrm{i}} H_{x,\mu} U_{x,\mu}\, ,
\end{equation}
where $H_{x,\mu}$ is the momentum conjugate to the field $ U_{x,\mu}$
and takes the values in su(2), the Lie algebra of SU(2).
$V_{x,\mu}$ denotes the staples around the link $U_{x,\mu}$, 
i.e.\ the incomplete plaquettes that arise in 
the differentiation\footnote{Details of the HMC algorithm for gauge theory
can be found in Ref.\ \cite{LIPPERT}.}. The classical
trajectories are computed using 
the leap-frog scheme, which consists of a sequence of 
intermediate points $(j=0,\dots,N_{\mathrm{MD}}-1)$ of the following form
\begin{eqnarray}
H_{x,\mu} (\Delta t' /2) &=& H_{x,\mu} (0) + 
\frac{\Delta t'}{2} \dot{H}_{x,\mu}(0) \, , \nonumber \\
U_{x,\mu} \left ( (j+1) \, \Delta t' \right ) 
&=& \exp \left ( {\mathrm{i}} \, \Delta t'
H_{x,\mu} \left ( j\, \Delta t' + \frac{\Delta t'}{2} \right ) \right )
U_{x,\mu} (j\, \Delta t') \nonumber \, , \\
H_{x,\mu}  \left ( j\, \Delta t' + \frac{\Delta t'}{2} \right ) &=& 
H_{x,\mu} \left ( j\, \Delta t' - \frac{\Delta t'}{2} \right ) 
+ \Delta t' \dot{H}_{x,\mu}
(j\, \Delta t') \, , \nonumber \\
H_{x,\mu} ( t' ) &=& 
H_{x,\mu} \left ( t' - \frac{\Delta t'}{2} \right ) 
+ \frac{\Delta t'}{2} \dot{H}_{x,\mu}
(t') \, .
\end{eqnarray}
The scheme is exact up to ${\cal O}({\Delta t '}^2)$. In order to generate the
desired Boltzmann distribution and to account for the discretization errors,
the new configuration is  only accepted with probability $P= \mbox{min}
\{ 1, \exp(-\Delta {\cal H}) \}$, where $\Delta {\cal H} = {\cal H} 
(U(t'),H(t')) -{\cal H}(U(0),H(0))$ and the Hamiltonian is given by
\begin{equation}
{\cal H}[U,H] = \frac{1}{2} \sum_{x,\mu} {\mathrm{Tr}}\,({H_{x,\mu}}^2) + S[U] \, .
\end{equation}

\section{Numerical results}

We perform simulations with $ L_{\mathrm{t}}=4$ and
$L_{\mathrm{s}}=8$, $16$ and $24$ at the
critical point $4/{g_{\mathrm{c}}}^2=2.2989$
and in the neighbourhood\footnote{For simulations in equilibrium these values 
would be considered to be far outside the critical region,
especially for $ L_{\mathrm{s}}=16$ and $24$. However, these values are 
close enough to the critical point to calculate the derivative of the 
magnetization with respect to $\tau$ in the short-time region.}
at $4/g^2=2.2689$ and $2.3289$.
Starting from the ordered initial
state, i.e.\ all link variables $U_{x,\mu}$ are set to unity,
we measure the magnetization $M$ and the cumulant $U$
as a function of the MC time $t$. 
The system is updated with the HMC algorithm, where we fixed
the length of the trajectory to $t'=0.32$. A unit in the MC time
$t$ is defined as one global Metropolis step. Simulations are
performed up to $t=400$ global MC steps.
The average is taken over ${\cal O}(1000)$ samples for $L_{\mathrm{s}}=8$,
${\cal O}(100)$ samples for $L_{\mathrm{s}}=16$ and
${\cal O}(10)$ samples for $L_{\mathrm{s}}=24$.
Statistical errors are calculated by dividing the data into different
subsamples. Systematic errors are estimated by the results of 
different system sizes and different time intervals, i.e.\
we examined the dependency of the critical exponents from the fitted interval
$t=[t_{\mathrm{min}},t_{\mathrm{max}}]$ and the space direction
$L_{\mathrm{s}}$. The quoted error is a sum of the statistical and systematic
error.

\begin{figure}
\begin{center}
\mbox{\epsfxsize=11.0cm
\epsfbox{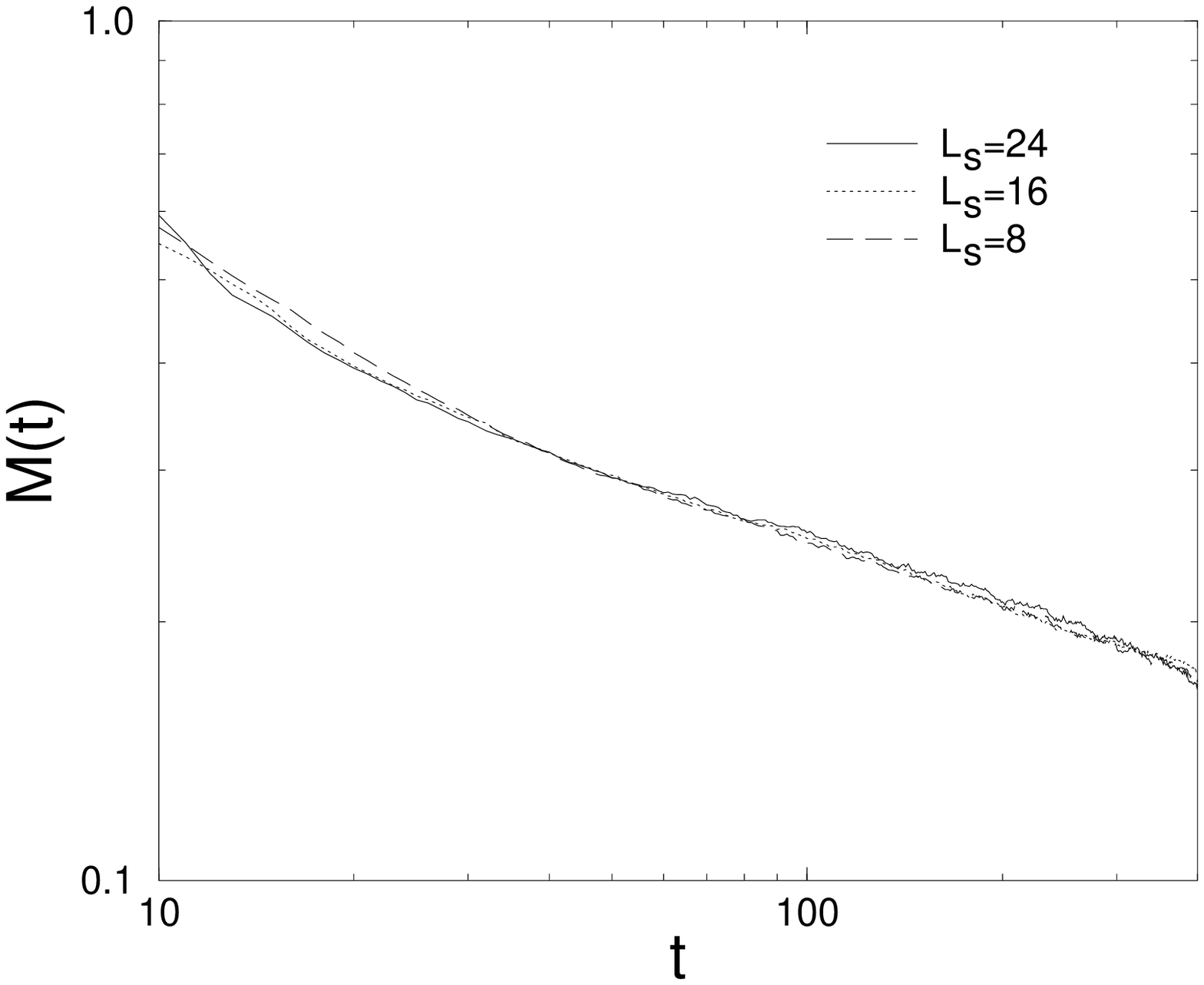}}
\end{center}
{\caption{\label{fig_magsize}
Time evolution of the magnetization at the critical point starting from the
ordered state for $L_{\mathrm{s}}=8$, $16$ and $24$.
The length of a trajectory of a HMC step was $t'=0.32$ with
$N_{\mathrm{MD}}=80$ intermediate points.
}}
\end{figure}

In Fig.\ \ref{fig_magsize} we plot the time evolution of the magnetization at
the critical point for different system sizes on a double
logarithmic scale. Statistical errors are of the order of the distance between the 
curves. For $L_{\mathrm{s}}=16$ a trajectory of the HMC algorithm
consists of $80$ steps with $\Delta t'=0.004$.
To get comparable results for the other lattices ($L_{\mathrm{s}}=8$, $24$), 
one can not perform 
simulations with the same parameters ($N_{\mathrm{MD}}$, 
$\Delta t'$)\footnote{In case of the local Metropolis algorithm
simulations of different system sizes can be simply performed with the same
parameters.}. The reason is the global accept/reject step. The difference
of the final and initial values of the Hamiltonian increases for larger lattices,
so that using the same parameters would result in a lower acceptance
rate. Therefore, we scale the step size with 
$\Delta t' \sim {L_{\mathrm{s}}}^{-1}$ to get comparable acceptance rates. 
Since the trajectory length $t'$ is constant, we have to scale 
$N_{\mathrm{MD}} \sim L_{\mathrm{s}}$ so that the CPU
time scales also with $L_{\mathrm{s}}$.
In all cases ($L_{\mathrm{s}}=8$, $16$, $24$) we get an acceptance rate of
approximately $99\%$.

However, also if we scale the parameters in the way described 
there are large deviations for the different system sizes 
at small times. This might arise
from statistical effects coming from the global accept/reject step.
Especially for the largest lattice we have a small statistics and the effect
of a global reject step at small times is high. 
Also, the acceptance rate (which is calculated by averaging over 
the 400 MC steps)  for early times is relatively small.
However, the results show that finite size effects coming from the finite
space dimension are small up to $t=400$, i.e.\ the influence of finite 
lattice size in Eq.\ (\ref{eqplm}) do not show up in the time
interval used\footnote{A system starts to show finite size
effects after a time scale proportional $L^z$, which leads
to deviations from the power law behaviour.}. 
Therefore, $L_{\mathrm{s}}=16$
is large enough to avoid systematic errors from too small lattices.

Obviously, the magnetization which is shown in Fig.\ \ref{fig_magsize}
can be described by a power law behaviour if we leave out the data 
for small times up to $t_{\mathrm{mic}}$.
Thus, the time interval shown is completly within the short-time regime.
The  microscopic time scale during which the non-universal 
behaviour is swept away is about $40$.
The slope of the curve yields the value of the exponent $c_1=\beta/\nu z$.
The results of the different space dimensions, subsamples and 
time intervals lead to $c_1=0.248(5)$.

\begin{figure}
\begin{center}
\mbox{\epsfxsize=11.0cm
\epsfbox{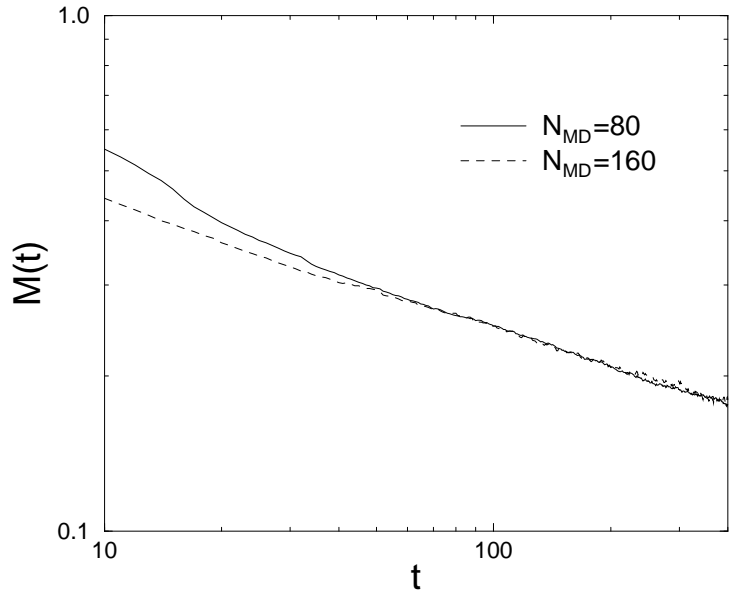}}
\end{center}
{\caption{\label{fig_Mtaus}
Magnetization as a function of time for $L_{\mathrm{s}}=16$
and $t'=0.32$ at the critical point. A trajectory consists
of $N_{\mathrm{MD}}=80$ steps in the first case (full line)
and $N_{\mathrm{MD}}=160$ steps in the second case (dashed line).
}}
\end{figure}

The critical exponent $c_1$ should be independent of the step size
$\Delta t'$ and number of steps $N_{\mathrm{MD}}$ of the trajectory.
However, a change of these parameters can lead to a different
microscopic time scale $t_{\mathrm{mic}}$ and a change of the statistical errors.
To examine the influence of the parameters $\Delta t'$ and
$N_{\mathrm{MD}}$ we study the short-time critical dynamics of the magnetization
with  $\Delta t'=0.002$ and $N_{\mathrm{MD}}=160$
for $L_{\mathrm{s}}=16$. The results are compared with the measurements 
using $\Delta t'=0.004$ and $N_{\mathrm{MD}}=80$
and are shown in Fig.\  \ref{fig_Mtaus}.
The acceptance rate changes from 98.7\% for $N_{\mathrm{MD}}=80$
to 99.7\% for $N_{\mathrm{MD}}=160$, while the CPU time increases 
by a factor of two. Although the changes for the acceptance rate are small, 
there are large deviations between both measurements
at small times. Figure \ref{fig_Mtaus} shows that the microscopic
time scale $t_{\mathrm{mic}}$  decreases for smaller $\Delta t'$.
The difference between both simulations at larger times are negligible and
the values for the critical exponent $c_1$ coincide within
statistical errors.

\begin{figure}
\begin{center}
\mbox{\epsfxsize=11.0cm
\epsfbox{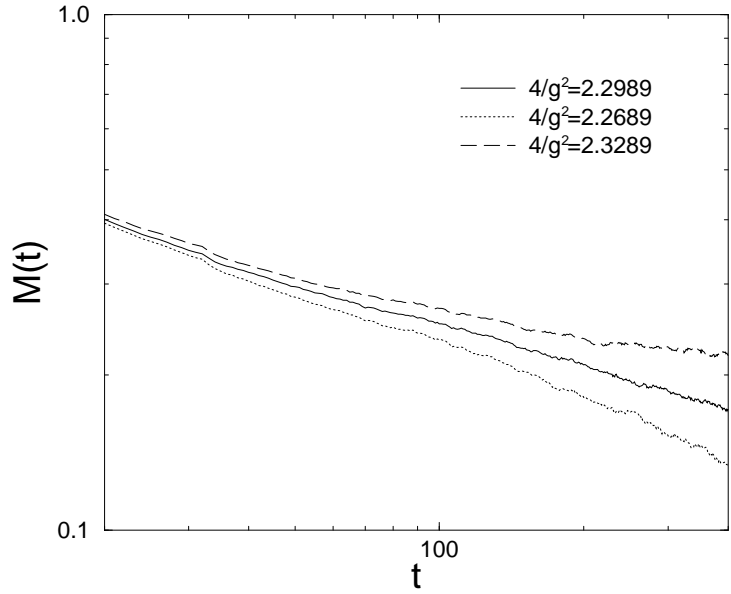}}
\end{center}
{\caption{\label{fig_Mpm}
Time evolution of the magnetization for three values of the coupling
constant with $L_{\mathrm{s}}=16$ and $N_{\mathrm{MD}}=80$. 
}}
\end{figure}

\begin{figure}
\begin{center}
\mbox{\epsfxsize=11.0cm
\epsfbox{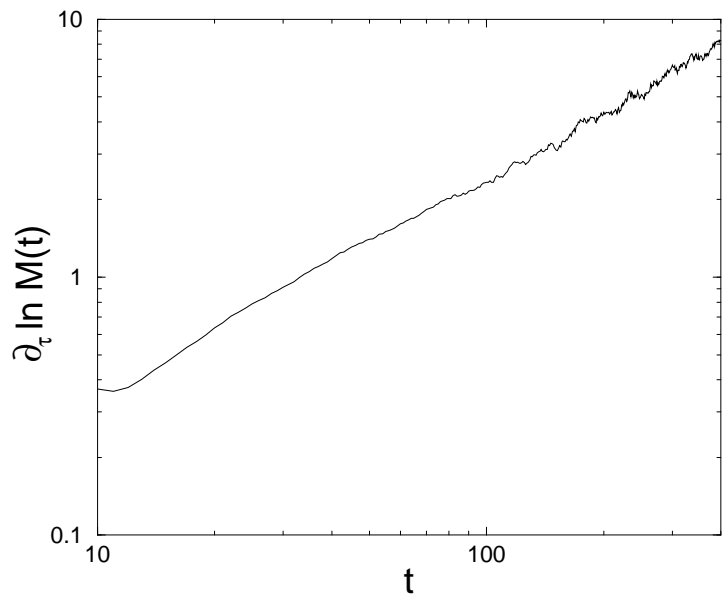}}
\end{center}
{\caption{\label{fig_derivM}
Logarithmic derivative  of the magnetization with respect
to $\tau$ taken at $g_{\mathrm{c}}$, obtained from the curves
shown in Fig\ 3.
}}
\end{figure}

To extract the critical exponent $c_{l1}$, we measure the magnetization
as a function of time also below and above the critical point. The simulations
are performed with $L_{\mathrm{s}}=16$, $N_{\mathrm{MD}}=80$
and $\Delta t'=0.004$. The results are visualized on log-log scale in Fig.\ 
\ref{fig_Mpm}. These data are used to calculate the logarithmic derivative 
of the magnetization with respect to $g_{\mathrm{c}}$. This is shown
in Fig.\ \ref{fig_derivM}. The slope provides 
$c_{l1}=0.83(3)$, where the error of 
$c_{l1}$ is dominated by systematic effects. 
In principle, one can also use the simulations to estimate the
critical point. This is done by searching the best power law behaviour of 
$M(t)$ between the two values $4/{g_1}^2=2.2689$ and $4/{g_2}^2=2.3289$ as
described in Ref.\ \cite{SCHZHE}. Namely, the best straight-line fit to curves obtained
by quadratic interpolation for $g_1> g > g_2 $ is sought. However,
our statistic is not good enough so that we do not perform 
this analysis.

The final step is to determine the critical exponent $z$. This is done
by measuring the cumulant $U(t)$ and extracting the exponent $d/z$.
Our simulations yield $z=2.0(1)$. Thus we get $\nu=0.60(5)$ and
$\beta=0.30(2)$. These results are in agreement (within
statistical errors) with those obtained from simulations
in equilibrium \cite{SU2EXP}, which are $\nu=0.630(11)$ 
and $\beta=0.328(6)$.
Also, the data coincide with those of the three-dimensional 
Ising model \cite{JAMASCZH}. Performing similar measurements 
(but using the Metropolis algorithm) one
gets in this case
$\nu=0.6327(20)$, $\beta=0.3273(17)$, and $z=2.042(6)$. 

The advantage of the
short-time dynamic approach compared to simulations in 
equilibrium is that is free of critical slowing
down since the spatial correlation length is small within the 
time regime, even at or near the critical point. Thus
in case of local algorithms (Metropolis, heatbath) the CPU
time to get comparable results for different systems
is independent of the system size, while simulations in the
long-time regime suffer from critical slowing down. 
However, if we use the 
HMC algorithm increasing the system size leads 
also for simulations in the short-time regime  to an
increase of the CPU time since we have to scale the
step size $\Delta t'$. Therefore, determing the static 
critical exponents can be done easier if we use
local algorithms (and short-time dynamics). 
However, if one wants to measure the 
dynamic critical exponent $z$ or include fermions 
in the simulation one has to use the HMC
algorithm. In this case using the dynamic relaxation
is advantageous compared to conventional simulations 
in equilibrium.

\section{Summary and conclusions}

We presented comprehensive Monte Carlo simulations of the short-time
critical dynamics for SU(2) lattice gauge theory in $(3+1)$ dimensions.
The dynamics of the system was given by the HMC algorithm.
Starting from the ordered state, the magnetization (Polyakov loop),
its derivative with respect to the coupling constant and the cumulant
were measured at the critical point. The observables obey a power
law behaviour after some microscopic time scale $t_{\mathrm{mic}}$
as expected. The critical exponents $\beta/\nu z $, $1/\nu z$
and $d/z$ are determined from these time dependencies.
The results support a universal short-time scaling behaviour
for SU(2) gauge theory in $(3+1)$ dimensions. 
The values for the static exponents $\beta$ and $\nu$ agree
within statistical errors with those measured in equilibrium and with those
of the three-dimensional Ising model.
Thus the $(3+1)$-dimensional SU(2) lattice gauge theory  
and the Ising model in three dimensions are in the same universality class.
The dynamic critical exponent for the HMC algorithm was determined with 
$z=2.0(1)$. The work could extended to SU(2)
lattice gauge theory with dynamical fermions.

\begin{ack}
Critical comments on our draft by Lothar Sch\"{u}lke are
gratefully acknowledged. 
Especially I benefitted from discussions with Inno Vista.
This work was supported in part
by the Deutsche Forschungsgemeinschaft under Grant No.\ DFG Schu 95/9-1.
\end{ack}

\newpage

\end{document}